\documentclass[prb,twocolumn,amsmath,amssymb,aps,superscriptaddress,eqsecnum]{revtex4-1}
\pdfoutput=1
\usepackage{graphicx}
\usepackage{dcolumn}
\usepackage{slashed}
\usepackage{epstopdf}
\usepackage{bm}
\begin{document}

\title{Interlayer coherent composite Fermi liquid in Half-filled Landau Level bilayers, a window toward the hidden $\pi$ Berry phase}

\author{Yizhi You}
\affiliation{Department of Physics and Institute for Condensed Matter Theory, University of Illinois at Urbana-Champaign, 
Illinois 61801}

\date{\today}
\begin{abstract}
For interacting 2D electrons in the presence of magnetic field at half filling, the system forms a `composite Fermi liquid(CFL)' with emergent Fermi surface and exhibits metallic behavior, based on the standard Halperin, Lee, and Read(HLR) field theory. Recently, Son introduces a composite Dirac theory as the low energy effective description of Half-filled Landau Level. Such theory exhibits particle-hole symmetry and the underlying composite Fermi surface displays a robust $\pi$ Berry phase. In this paper, we start from the bilayer Half-filled Landau Level system where the two composite Fermi surface acquires interlayer coherence and forms bonding/anti-bonding composite fermi sea. The corresponding interlayer coherent composite Fermi liquid(ICCFL) phase provides a straightforward landscape to verify the Dirac nature in Son's theory and extract the hidden Berry phase structure of the composite Fermi surface.  The ICCFL phase contains two Fermi surfaces which are detached in most regions but adhesive at two hot spots.  Such nematic structure is a consequence of the Berry phase encoded in the Dirac Fermi surface which is absent in HLR theory. Due to the nematicity in ICCFL, the system supports half-quantum vortex with deconfined $\frac{\pi}{2}$ gauge flux and the phase transition toward ICCFL contains a Lifshitz criticality with $z=3$ dynamical exponent.
In addition, the exciton order parameter carries topological spin number so the ICCFL contains a unique Wen-Zee term which connects EM response with the background geometry curvature.
\end{abstract}

\maketitle

\section{Introduction and Motivation}
Interacting 2d electrons in the presence of strong magnetic field exhibit rich phase diagram and exotic phenomenon\cite{willett1987observation}. Over the past decades, much had been explored in such platform and a variety of theoretical frameworks had been done to characterize different phases. Apart from the incompressible fractional quantum Hall(FQH) states, there also exist a class of compressible quantum liquid state when electrons are at even filling factor $\nu=\frac{1}{2n}$\cite{willett1987observation,halperin1993theory,Pasquier-1998}. Theoretical framework on such compressible quantum liquid state was first developed by Halperin, Lee, and Read(HLR)\cite{halperin1993theory}, based on the framework of composite fermion theory with flux attachment\cite{Jain1989,Lopez-1991}. When each fermion was attached with $2n$ flux, the composite fermion does not perceive the magnetic flux and they eventually form a Fermi surface with strong interaction mediated by dynamical gauge fields. The flux attachment generates a Chern-Simons gauge field and the gauge fluctuation was damped by the gapless composite Fermi(CF) surface. Meanwhile, the gauge fluctuations also soften the Fermi surface and the quasiparticles acquire finite lifetime\cite{Polchinski-1993,Stern-1999,Nayak-1994,murthy2016nu,Mross-2010}.

The HLR theory displays a clear picture for compressible quantum Hall phase. However, once we turn away the Landau Level(LL) mixing and project the physical degree of freedom into the lowest LL, the HLR theory cannot reproduce the particle-hole(PH) symmetry for half-filled Landau Level\cite{Kivelson-1997,balram2016nature,levin2016particle,Barkeshli-2015,Lee-2007,Levin-2007,Girvin-1984,wang2016particle,cheung2016weiss,murthy2016nu,lee1999unsettled}. Recently, Son proposed a new composite Dirac liquid theory for Half-filled LL system\cite{Son-2015}, where the Fermi surface can be characterized as the gapless Dirac fermion at finite chemical potential with neutral charge. This theory reproduce the exact Hall conductivity $\sigma_{xy}=\frac{e^2}{2h}$ and solve the ambiguity of Pfaffian/anti-Pfaffian states related by particle-hole conjugation\cite{lee2007particle,levin2007particle,Kivelson-1997}. 
Afterwards, a group of pioneers\cite{seiberg2016duality,wang2016composite,mross2016explicit,wang2016half,Kachru-2015,Metlitski-2015,mulligan2016particle,potter2016realizing,mulligan2016emergent} develop similar composite Dirac liquid theory in a microscopic point of view and a variety of numerical measurements\cite{geraedts2016half} and experimental proposals\cite{cheung2016weiss,potter2016thermoelectric} had been raised to verify this theory. A numerical study on the absence of $2k_f$ singularity in PH even channel provides a smoking gun for the presence of an emergent Fermi surface with Dirac structure\cite{geraedts2016half}. In addition, the thermoelectric transport measurement proposal indicates the charge neutrality of the Fermi surface could be demonstrated from the off-diagonal thermopower\cite{potter2016thermoelectric}.

The main difference between Son's composite Dirac liquid(CDL) theory and the conventional HLR theory comes from the $\pi$ Berry phase encoded in the composite Fermi surface. In this paper, we intend to provide a new sight to justify the emergent $\pi$ Berry phase in the half-filled Landau level system.  In both theory, fermion degree of freedom couples to a dynamical U(1) gauge field. The Fermi surface thereby becomes strongly interacting and the probe of the concealed $\pi$ Berry phase in such non-Fermi liquid can be rather challenging. In addition, due to the charge neutrality in Son's CDL, one cannot verify the Berry phase via the Shubnikov-de Haas oscillation measurement\cite{wang2016half,wang2016composite}.

In order to reveal the hidden $\pi$ Berry phase\cite{chen2016berry,haldane2004berry}, one has to look into some physical quantity which is sensible to the Berry phase of the Fermi surface.
Just like the Fu-Kane superconductor\cite{fu2008superconducting}, the s-wave pairing of the Dirac Fermi surface in Son's theory finally give rise to $p+ip$ SC due to the Berry phase\cite{wang2016half,wang2016composite}. However, for the HLR theory, a conventional Fermi surface with polarized spin also encounter with p-wave pairing instability so one cannot distinguished Son from HLR via pairing channels\cite{Read-1998,read2000paired}. 

For a half-filled Landau level bilayer, when two layers are at intermediate distance, the two composite Fermi surfaces form an interlayer coherent composite Fermi liquid(ICCFL)\cite{sodemann2016composite,alicea2009interlayer,cipri2014gauge}, which could be expressed as the exciton condensate of the composite fermion in the particle-hole channel between two layers. For a conventional Fermi surface with no berry phase (HLR), the interlayer coherent state split the degeneracy of the two Fermi surfaces and the fermions form two isotropic Fermi surfaces with different wave vectors. Meanwhile, for a bilayer system with two Dirac Fermi surface on each layer (Son's theory), the exciton condensation, in some specific channel, has $p+ip$ symmetry in the order parameter as a consequence of the Berry phase.  As a result, the exciton order parameter carries topological spin and one can measure them in terms of the Wen-Zee effect where electromagnetic response intertwined with geometry curvature. In addition, in the PH odd channel, the exciton order parameter contains nodal structure due to the $\pi$ Berry phase and forms a spontaneous nematic state. Such anisotropy could be measured in terms of density susceptibility in the static limit. Owing to the nematicity in ICCFL, the system supports half-quantum vortex defect trapping $\pi/2$ gauge flux. The Goldstone boson would be overdamped in the nematic ICCFL phase as a consequence of the Dirac structure on the composite Fermi surface and the phase transition toward the nematic ICCFL is a Lifshitz criticality with $z=3$ dynamical exponent.

In the rest part of this paper, we would systematically investigate the bilayer half-filled Landau level system, where each layer contains 2d interacting electrons at $1/2$ filling. When the composite fermions contain repulsive interlayer interaction, the composite Fermi surface on the two layers breaks interlayer U(1) symmetry and forms interlayer coherent composite Fermi liquid(ICCFL). We would compare the behavior and character of ICCFL based on HLR and Son theory. Our result would demonstrate that the hidden $\pi$ Berry phase in Son's CDL could induce nematicity and Wen-Zee effect in the ICCFL phase and thus provides a feasible way to test and verify the Dirac nature of the half-filled Landau level. Possible experimental implication and measurement proposal would be discussed in details.

\section{Interlayer coherence in composite fermi liquid}

In bilayer half-filled Landau Levels,  the system could encounter a rich class of instabilities and therefore form new states of matter\cite{sodemann2016composite,alicea2009interlayer,eisenstein2014exciton}. At small interlayer distance, the electrons on different layers acquire coherence and the spontaneous tunneling between layers leads to the interlayer superfluid state $\langle c^{\dagger}_1 c_2\rangle \neq 0$\cite{sodemann2016composite,alicea2009interlayer,eisenstein2014exciton}. Such phase involves condensation of bound state between the physical electrons(filled Landau Levels) in one layer and holes (empty Landau Levels) in the other layer. The resultant state could be regarded as bilayer (111) state with counterflow superfluidity. In the literature of Son's Dirac fermi liquid, the bilayer pairing of the composite Dirac liquid, enhanced by the gauge fluctuation\cite{sodemann2016composite}, is equivalent to the interlayer superfluid state which could be verified by the dual picture. 

When the half-filled Landau Level bilayers are separated at an intermediate distance, the system demonstrates tremendous rich physics which was studied by experiments and numerical simulations\cite{sodemann2016composite,alicea2009interlayer,milovanovic2016pairing,milovanovic2015meron}. One of the prominent exotic phase raised by Alicea $ et$ $al.$\cite{alicea2009interlayer} suggests there could appear a new composite Fermi liquid where the composite fermions acquire interlayer coherence but the electrons do not. Such phase contains two Fermi surfaces with different Fermi wave vectors splitted by interlayer coherence. This state is compressible with respect to symmetric currents but contains quantized Hall response in the counterflow channel. Below we would briefly review the theory of interlayer coherent composite Fermi liquid phase from HLR theory.

\subsection{Interlayer coherence in HLR theory}
In HLR theory, the fermions in half-filled Landau Level is attached with two flux so the composite fermions form a Fermi surface with emergent gauge field containing a Chern-Simons term. Alternatively, such composite Fermi surface could also be reached via slave particle theory by taking $\Psi=\Psi^{cf} b$. Here $\Psi$ is the physical electron carrying EM charge while $\Psi^{cf}$ is the composite fermion carry gauge charge $a$. $b$ is a slave boson carrying both gauge charge and EM charge. When boson is at half-filling with respect to the magnetic field, the boson forms FQH state with $ \frac{1}{8\pi}(A-a) \wedge d (A-a)$ Hall response. Thus the bilayer theory for half-filled Landau Level is,
\begin{align} 
&\mathcal{L}={\Psi}^{\dagger}_{cf,i}(iD_0+\frac{(D^2_{x}+D^2_{y})}{m})\Psi_{cf,i}\nonumber\\
&+\frac{1}{8\pi} (A_i-a_i) \wedge d (A_i-a_i) +..\nonumber\\
&D_{\mu}=\partial_{\mu}+ia_{\mu}
\label{hlr}
\end{align}
Here $i$ refers to layer index.

The interlayer coherent composite Fermi liquid state refers to the coherence of the composite fermion $\langle {\Psi}^{\dagger}_{cf,1} \Psi_{cf,2} \rangle \neq 0$ which spontaneously breaks the interlayer U(1) as of $a^-=(a_1-a_2)/2$. Such state would split the original degenerate composite Fermi surface into two surfaces with different wave vectors. However, even the theory contains gapless fermions, the quantum Hall response in the counterflow channel $A^-=(A_1-A_2)/2$ is still nonzero. Apart from the transport signature in the counterflow channel, one can also probe this phase via static density-density correlations. A normal Fermi surface contains $2k_f$ singularity in static density-density correlation. Interlayer coherence splits the two Fermi surface with enlarged/shrunk wave vectors $k_f\pm a$ ($a$ depends on the $\langle {\Psi}^{\dagger}_{cf,1} \Psi_{cf,2} \rangle$ parameter). The resultant Lindhard function contains four singularities at $2k_f, 2a, 2(k_f \pm a)$, which refers to the distance between Fermi surfaces. The calculation detail of Lindhard function of ICCFL phase would be discussed in appendix A.

The transition from interlayer coherent composite Fermi Liquid phase toward the interlayer superfluid (111) state involves the interlayer coherence of bosons. When $\langle b^{\dagger}_1 b_2\rangle \neq 0$, both composite fermion and slave boson acquires interlayer coherence, bring about the interlayer superfluid state $\langle c^{\dagger}_1 c_2\rangle \neq 0$. If  $\langle b^{\dagger}_1 b_2\rangle = 0$, the interlayer coherence only appears in composite fermion level. Thus, the condensation of 
$\langle b^{\dagger}_1 b_2\rangle $ drives the phase transition from interlayer coherent composite Fermi Liquid toward interlayer superfluid state\cite{alicea2009interlayer}. The transition theory could be described by the superfluid boson coupling with a dynamical compact U(1) gauge theory. Such transition is beyond conventional Landau paradigm as the boson here is a fractionalized degree of freedom\cite{senthil2008critical}.

\subsection{interlayer coherence in CDL theory}
In Son's CDL theory, the composite Fermi surface, with the structure of Dirac Fermion is charge neutral while the gauge charge carried by the composite fermion couples to the electromagnetic field. 
\begin{align} 
&\mathcal{L}=\bar{\Psi}^D \slashed{D}_{\mu} \Psi^D-\frac{1}{4\pi} A da +\frac{1}{8\pi} A dA\nonumber\\
&D_{\mu}=i\partial_{\mu}+a_{\mu}
\label{son}
\end{align}
The PH symmetry here is anti-unitary and acts in a similar way as the usual time reversal, 
\begin{align} 
\mathcal{CT} : & \Psi^D \rightarrow i\sigma_y \Psi^D ,\nonumber\\
& a_0 \rightarrow a_0,\nonumber\\
& a_x,a_y \rightarrow -a_x,-a_y
\end{align}
The Dirac Fermi surface structure is a consequence of Landau Level projection and PH symmetry. Before LL projection, the fermion bound with two vortices(correlation holes) form a composite Fermi surface at charge neutrality. LL projection shifts one vortex away from the fermion center while the shifted direction is orthogonal to the Fermi momentum of the composite fermion on the Fermi surface\cite{wang2016half}. Consequently, the vortex-fermion bound state with a shift forms a charge dipole and it carries momentum perpendicular to the dipole direction. This is similar to the spin-orbital coupling in Dirac fermions where the spin and momentum are locked\cite{wang2016half}. In addition, when we go around the Fermi surface, the fermion's momentum angle winds around $2\pi$ and so is the dipole. As the dipole's self-rotation accumulates a $\pi$ Berry phase, the Fermi surface also carries $\pi$ berry phase which exactly matches the theory of Dirac Fermi surface. The PH symmetry($\mathcal{CT}$) rotates the dipole by 180 degrees so one can express the symmetry operator as $ i\sigma_y$ which rotates the Dirac spinor. The mass term of the Dirac fermion is absent here as this would break PH symmetry and destroy the locking between charge dipole and momentum.

Starting from the bilayer half-filled LL system, when the distance between the two layer goes small and interlayer interaction becomes crucial, there appears a variety of exotic phases. At this stage, we focus on the ICCFL phase involving condensation of composite fermion in particle-hole channel on different layers $\Psi^{\dagger D}_1\Psi^{D}_2$. Such condensation turn on the interlayer coherence between the two composite Dirac liquid and break the relative U(1) symmetry between two layers. Just as the Cooper pairs can inherit the Berry phase structure of the original Fermi surface\cite{li2015topological,wang2016topological,wang2016composite},
when coherence happens between states on two disjoint Fermi surface with Berry phase, the exciton acquires
nontrivial Berry structure from the underlying single-particle Fermi surfaces. Hence, the hidden Berry structure of the CDL would be revealed in the interlayer coherent composite Dirac liquid state.

We can reach the full theory of the bilayer system as,
\begin{align} 
&\mathcal{L}=\mathcal{L}_f+\mathcal{L}_{\phi} \nonumber\\
&\mathcal{L}_f= \bar{\Psi}^D_i (i  \slashed{\partial}_{\mu}+ \slashed{a}_{\mu,i}) \Psi^D_i-\frac{1}{4\pi} A_i da_i +\frac{1}{8\pi} A_i dA_i \nonumber\\
&+\phi_n \Psi^{\dagger D}_{i} \sigma_n \Psi^{D}_{j}+\phi_0 \Psi^{\dagger D}_{i}  \Psi^{D}_{j}+h.c.+.. \nonumber\\
 &\mathcal{L}_{\phi}=|(i\partial_{\mu}+2a^-_{\mu})\phi|^2-s|\phi|^2-r/2 |\phi|^4+....
 \nonumber\\
&a^+=\frac{a_1+a_2}{2},a^-=\frac{a_1-a_2}{2},
\label{action}
\end{align}
Here $\phi$ is the exciton order parameter which is spontaneously generated by interaction to induced interlayer coherence. $i,j$ refers to layer index.

\subsubsection{Nematic ICCFL}
\begin{figure}[h!]
 \centering
  \includegraphics[width=0.45\textwidth]{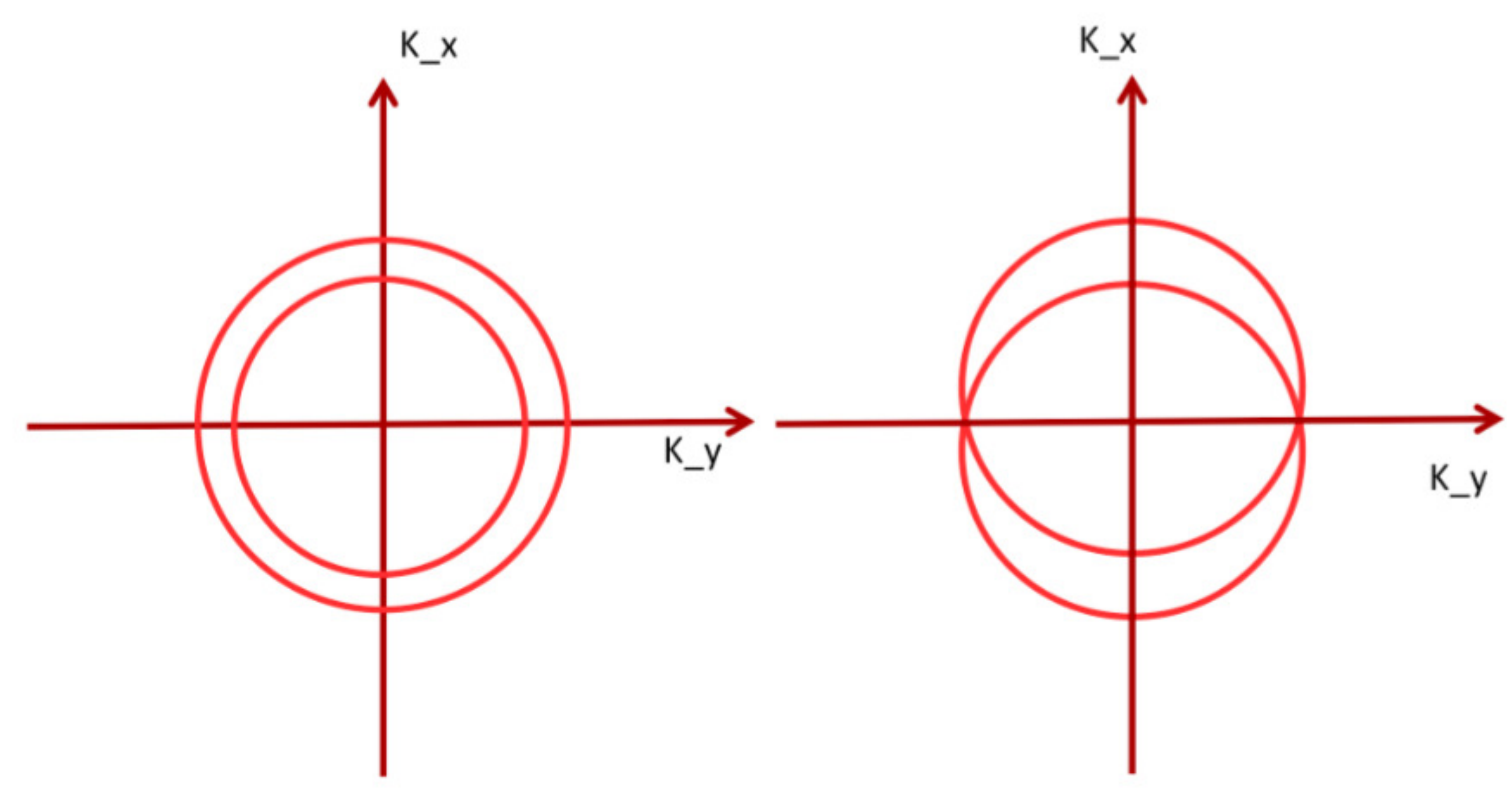}
 \caption{In the presence of interlayer coherence between composite fermions,  Left: the composite Fermi surface in the HLR theory. Right: the composite Fermi surface in Son's theory.}
 \label{exciton1}
 \end{figure}

There could be four independent interlayer coherent states,
\begin{align} 
& \phi=\phi^{1}+i \phi^{2} \nonumber\\
& \phi_{a}^1 \rightarrow \langle \Psi^{\dagger D} \sigma_x \tau_x \Psi^{D}\rangle, \phi_{a}^2\rightarrow  \langle \Psi^{\dagger D} \sigma_x \tau_y \Psi^{D}\rangle \nonumber\\
& \phi_{b}^1 \rightarrow  \langle\Psi^{\dagger D} \sigma_y \tau_x \Psi^{D}\rangle, \phi_{b}^2\rightarrow  \langle\Psi^{\dagger D} \sigma_y \tau_y \Psi^{D}\rangle \nonumber\\
&\phi_{c}^1\rightarrow \langle \Psi^{\dagger D} \sigma_z \tau_x \Psi^{D}\rangle, \phi_{c}^2\rightarrow  \langle\Psi^{\dagger D} \sigma_z \tau_y \Psi^{D}\rangle \nonumber\\
&\phi_{d}^1\rightarrow  \langle \Psi^{\dagger D} I \tau_x \Psi^{D}\rangle , \phi_{d}^2\rightarrow  \langle\Psi^{\dagger D} I  \tau_y \Psi^{D} \rangle
\label{ph}
\end{align}
Here $\sigma$ acts on Dirac spinor index while $\tau$ acts on layer index. 
One can defined the PH symmetry operator $\mathcal{CT}: \mathcal{K} i\sigma_y$, spatial Parity operator $\mathcal{CP}: \sigma_x$, and layer switching symmetry operator $\mathcal{X}: \tau_x$. All four exciton channels transform under $\mathcal{X}$ as $\phi \rightarrow \phi ^*$\cite{aaa}. $\phi_{a}$ is even under $\mathcal{CP}$ while $\phi_{b}$ is odd.
Only $\phi_{d}$ is PH even while else are PH odd. Here we only focus on the PH odd channel which inherits the Berry phase structure of the Dirac Fermi surface\cite{bbb}. As the theory only involve CF near the Fermi surface, one can project away the lower band and the fermions on the Fermi surface is written as $c_i=\psi_{\uparrow}+e^{i \theta_p} \psi_{\downarrow}$. Consequently, the projected exciton order parameter is,
\begin{align} 
& \phi_{a}\rightarrow  \langle c^{\dagger }_1 c_2 e^{i \theta_p} +c^{\dagger }_1 c_2 e^{-i \theta_p}  \rangle \sim \langle c^{\dagger }_1 c_2\rangle  \cos(\theta_p)  \nonumber\\
& \phi_{b} \rightarrow  \langle c^{\dagger }_1 c_2 e^{i \theta_p} -c^{\dagger }_1 c_2 e^{-i \theta_p} \rangle \sim \langle c^{\dagger }_1 c_2  \rangle \sin(\theta_p)   \nonumber\\
& \phi_{c}= 0  \nonumber\\
& \phi_{d}\rightarrow  \langle c^{\dagger }_1 c_2  \rangle
\end{align}
It turns out that the PH odd exciton order parameter is dressed with a phase factor depending on the momentum angle of the wave vector and such dependence is a consequence of the internal Berry phase structure carried by each Fermi surface. For exciton in the PH odd channel, the nonzero exciton $\phi_{a} \sim \cos(\theta_p)$($\mathcal{CP}$ even) or $\phi_{b} \sim \sin(\theta_p)$($\mathcal{CP}$ odd) contains nodal configuration where at two hot spots the exciton order parameter is zero\cite{bbb}. As a result, the Fermi surface splitting is non-uniform in momentum space. The Fermi surface was detached at most region but at the nodal points they adhere each other, as is shown in Fig \ref{exciton1}.  Such anisotropy is merely a consequence of the Berry phase structure in the underlying Dirac Fermi surface and is not expected in HLR theory.
 
The phase transition theory $\mathcal{L}_{\phi}$ in Eq.[\ref{action}] for nematic ICCFL shall carry additional damping term $\gamma \frac{i\Omega}{q} |\phi|^2$. This overdamped exciton mode change the scaling dimension at the quantum critical point with dynamical exponent $z=3$\cite{ruhman2014ferromagnetic,bahri2015stable}.  In the nematic phase, we have two transverse Goldstone modes with respect to the interlayer U(1) and rotation symmetry breaking. The Goldstone mode for interlayer U(1) would be gapped by the Higgs mechanism for $a^-$. The transverse mode for nematic fluctuation remains gapless and the nematic Goldstone boson is overdamped at ICCFL phase. The origin of such Landau damping effect of Goldstone boson at the symmetry breaking region could be traced back to the spin orbital coupled structure of the Dirac fermi surface.  When the system has spontaneous symmetry breaking, the survival of an overdamped Goldstone mode depends on the commutation relation between symmetry generators that are spontaneously broken and the conserved momentum of translation symmetry\cite{watanabe2014criterion}. The nematic ICCFL with respect to exciton order parameter $\phi_a, \phi_b$ in Eq.[\ref{ph}] breaks the rotation invariance(with symmetry generator $Q=\sigma_z$) of the Dirac spinor and turn on `spontaneous easy-plane magnetization'. Meanwhile, the momentum operator for Dirac Fermi surface is locked with the spin operator $\hat{P}_x=p_x\sigma_x,\hat{P}_y=p_y\sigma_y$. As the momentum operator $\hat{P}_x,\hat{P}_y$ does not commute with $Q$\cite{ruhman2014ferromagnetic,bahri2015stable}, the overdamped Goldstone mode disperse as $\omega \sim q^3$ which would enhance the non-Fermi liquid behavior of the Fermion in ICCFL.
This result is a consequence of the Dirac nature on the Fermi surface with momentum-spin locking and is not expected in HLR theory.

In addition, when the exciton falls into the PH odd channel as $\phi_{a} \sim \cos(\theta_k)$ which breaks rotation invariance, there exist a deconfined half-quantum vortex of the exciton order parameter (bound with $\pi$ disclination). This phenomenon is similar to the half-quantum vortex(HQV) in pair density waves\cite{gopalakrishnan2013disclination,berg2009charge,you2016geometry} where a half SC-vortex is bound to dislocation. Ergo, the minimal deconfined gauge flux for $a^-$ is $\pi/2$.  In our half-filled LL theory, a gauge flux of $2\pi$ implies an electron layer-charge-imbalance $N_{-}=1$ under $A^-=(A_1-A_2)/2$.
One experiment implication on this HQV is the layer imbalanced electron density in the presence of disorder.
Provided the system contains spatial disorders, $e.g.$ a $\pi$ disclination, the disclination would bound to the HQV and hence the electron density difference between layers with $N_{-}=1/4$  is observed in the disclination core.

The fermion theory of the exciton condensate phase after Fermi surface projection is,
\begin{align} 
 &\mathcal{L}_{f}= \phi(p) c^{\dagger }_1(p) c_2(p)+h.c.\nonumber\\
 &+(|p|-k_f ) (c^{\dagger }_1(p) c_1(p)+c^{\dagger }_2(p) c_2(p))
\end{align}

For exciton in the PH odd channel as $\phi(p)_a\sim \cos(\theta_p)$, the splitting distance between two Fermi surfaces varies in momentum space. At the nodal spot $\cos(\theta_p)=0$, the two Fermi surfaces are touched at hot spot while in other region, the two Fermi surface are detached. This structure could be detected in experiment(or numerics) in terms of static density-density correlation. The singularity of the Lindhard function contains four singularity at $2k_f, 2a\cos(\theta_p), 2(k_f \pm a\cos(\theta_p))$. At the nodal point when $\cos(\theta_p)=0$, these four singularity merged into one singularity at $2k_f$.

Such Fermi surface structure is similar to the $\alpha$ phase studied by Wu\cite{wu2007fermi}  $et$ $al.$ where the Fermi liquid instability happens in spin channel and the Fermi surface exhibit spontaneous anisotropic distortions due to electron dipole interaction in $l=1$ channel.

In our ICCFL phase, in order to induce interlayer coherence between two composite Dirac liquid, the interlayer repulsive interaction between composite fermion is essential.  One can write down the interlayer interaction in a general form,
\begin{align} 
&V_{int}= ~ \int d\bm{q} ~d\bm{k} ~d\bm{k'}\nonumber\\
&-V_1(q)  \Psi^{\dagger D}_{1}(k+q) \sigma_i \Psi^{D}_{1}(k)\Psi^{\dagger D}_{2}(k'-q) \sigma_i \Psi^{D}_{2}(k')\nonumber\\
&-V_2(q) \Psi^{\dagger D}_{1}(k+q)  \Psi^{D}_{1}(k) \Psi^{\dagger D}_{2}(k'-q)  \Psi^{D}_{2}(k')
\label{int}
\end{align}
 $V_1,V_2$ are the interaction potentials. Both interactions are PH even and invariant under orbital angular momentum($l=0$). However, the first term in Eq.[\ref{int}] involves interaction in spin channel.  For fermions near the composite Fermi surface, the Dirac structure locks the spin(dipole orientation) with the momentum. Consequently, any interaction in spin channel would effective produce interaction in orbital angular momentum channel. To demonstrate, we first project the fermions near Fermi surface as $c_i=\psi_{\uparrow}+e^{i \theta_p} \psi_{\downarrow}$. The interaction in Eq.[\ref{int}] becomes,
\begin{align} 
&V_{int}= ~ \int d\bm{q} ~d\bm{k} ~d\bm{k'}\nonumber\\
&-V_1(q)  \cos(\theta_k-\theta_{k'}) c^{\dagger}_{1}(k+q)  c_{1}(k) c^{\dagger}_{2}(k'-q) c_{2}(k')]\nonumber\\
&-V_2(q) c^{\dagger}_{1}(k+q)  c_{1}(k) c^{\dagger}_{2}(k'-q) c_{2}(k')
\label{inteff}
\end{align}
Here we assume the interaction potentials are short ranged and we take the long wavelength limit. The first term in Eq.[\ref{inteff}] effectively generates a dipole-dipole interaction in the $l=1$ orbital angular momentum channel. Such interaction channel favors a distorted exciton order parameter(which break rotation symmetry) as $\phi_{a}\sim \cos(\theta_p),\phi_{b}\sim \sin(\theta_p)$ in Eq.[\ref{ph}].  Meanwhile, The second term in Eq.[\ref{inteff}] is invariant under both orbital and spin angular momentum. Thus, such interaction channel favors an isotropic exciton order parameter $\phi_{d}$ in Eq.[\ref{ph}].

In half-filled Landau Levels, the interlayer interaction between composite fermion in dipole channel($l=1$) could arise naturally as a consequence of LL projection. The LL projection shifts one correlation hole(vortex) away from the composite fermions center and the corresponding CF has the form of a charged dipole with opposite charge($\pm e/2$) on its ends. Accordingly, the interaction between the CF on two layers has the form of dipolar interaction $V\sim \cos(\theta_k-\theta_{k'})$ which depend on their dipole orientation and thus affected by the momentum angle. ( $\theta_k-\theta_{k'}$ is the azimuthal angle between two CF dipole.)

Due to the presence of gauge fluctuation for $a^-$(which flows to finite strength under RG), the interlayer coherence is suppressed so a finite interaction strength is required\cite{sodemann2016composite,cipri2014gauge,zhuzheng}. In addition, to systematically determine which exciton channel in Eq. (\ref{spin}) is favored, one has to collect the full information of the composite fermion below the Fermi surface in order to calculate the free energy and susceptibility with respect to different channels. However, in Son's CDL, only the composite fermion near the Fermi surface is equal to the Dirac Fermi surface. The energy dispersion of composite fermion far below the chemical potential is still missing. Hence, at this point, we cannot determine which exciton channel is energetically favored.

The interlayer coherence between two Fermi surface leads to the interlayer drag conductivity which is measurable in experiment. Imagine we add an external electric field $E_x$ in the first layer, based on Eq.[\ref{son}], the electric field $E_x$ is bound to the composite fermion current $J^{D}_y$ so we expect there is a neutral composite fermion current $J^{D}_y$ in the first layer.  As the exciton condensation involves with the interlayer coherence between the composite particle and composite hole in different layers, an opposite composite fermion current $J^{D}_y$ in the second layer would appear meanwhile. Consequently, the electric field $-E_x$ on the second layer is induced.
 \begin{figure}[h!]
 \label{link}
 \centering
  \includegraphics[width=0.3\textwidth]{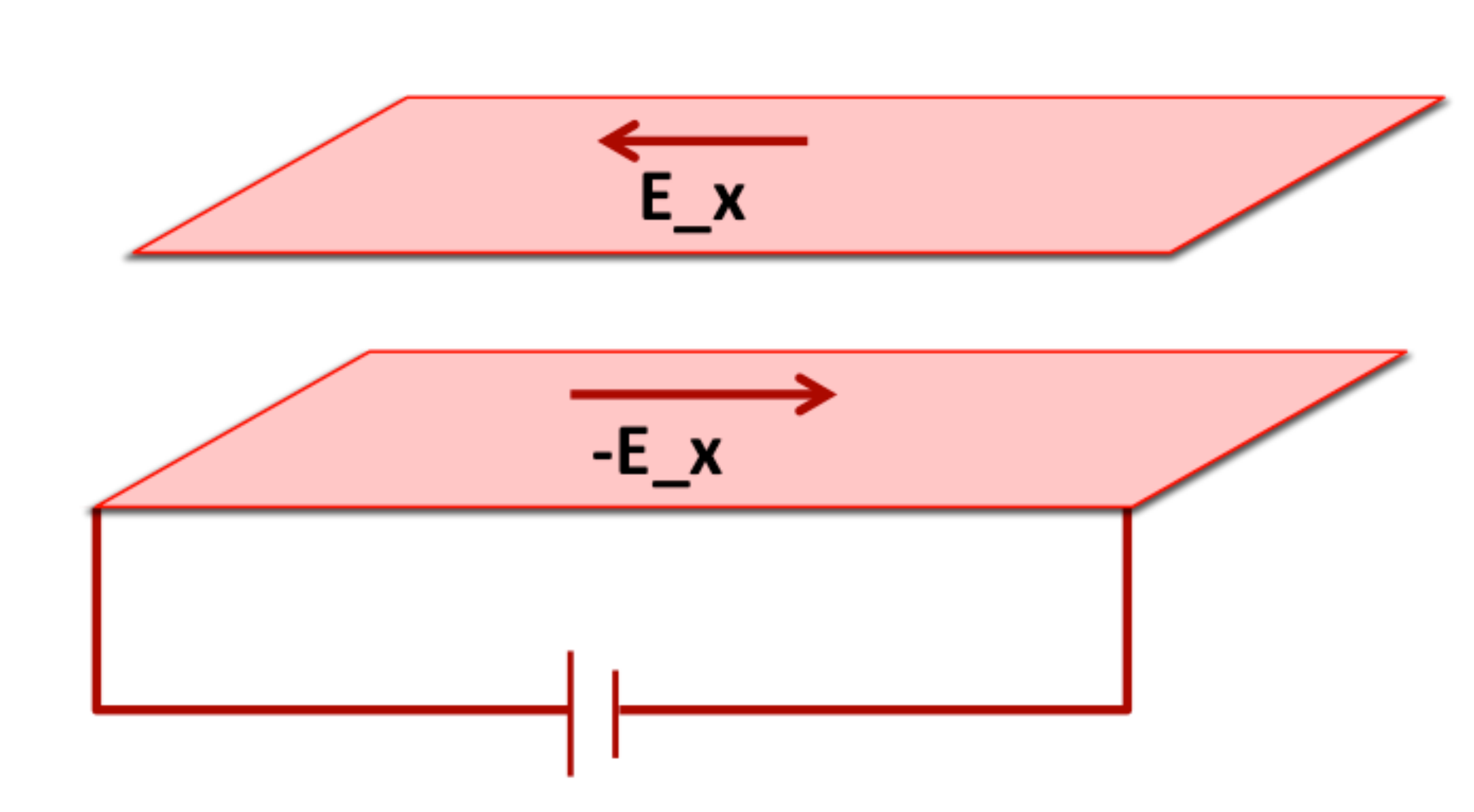}
 \caption{Placing an electric field in the first layer would induced an electric field on the second layer in the opposite direction.}
 \label{drag}
 \end{figure}
 Another measurable consequence of the ICCFL is the finite stiffness due to exciton condensation. The stiffness in the ICCFL could be measured in numerics by adding a twist boundary phase and analyze the energy cost by rotating the exciton order parameter of ICCFL with a small phase angle\cite{zhuzheng}.
 
However, in experimental measurement, it is hard to tell or control whether the exciton $\phi$ falls into the $\mathcal{CT}$ and $\mathcal{CP}$ odd/even channel. For $\mathcal{CT}$ even channel exciton field, the order parameter could be isotropic which make it impossible to tell the CDL with HLR theory. To solve this issue, we would now classify different excitons in terms of topological spin numbers. If the exciton carries a nontrivial topological spin number, one expects there should be some nontrivial geometric(or viscoelastic) response which is measurable in experiments.

\subsubsection{Wen-Zee effect}

For Dirac fermions, the Dirac spinor carries topological spin. Once we couple them with background geometry metric, the fermions acquires geometric phase with respect to the spin connection\cite{Wen_Zee,Cho-2014}.
\begin{align} 
& \psi_{\uparrow} \rightarrow \exp(i/2 \int \omega^{xy}_i dx_i)\psi_{\uparrow} \nonumber\\
& \psi_{\downarrow} \rightarrow \exp(-i/2 \int \omega^{xy}_i dx_i)\psi_{\downarrow}
\end{align}
$\omega^{xy}$ is the spin connection on the $x-y$ plane.
To comprehend the origin of this Berry phase induced by geometry curvature, we shall look back into the dipole picture of the composite surface after LL projection. The LL projection projects one vortex away from the fermion and the composite fermion forms a charge dipole whose direction is perpendicular to the momentum. When we place such dipole on non-flat geometry and do parallel transport, the dipole rotates with the local frame along the curve. This procedure accumulates Berry phase from dipole(Dirac spinor) rotation. Consequently, the spin connection would appear in the covariant derivative of the dipole(Dirac spinor)'s equation of motion.

As long as we define the topological spin for the Dirac fermion, one can also classify different exciton in terms of topological spins.
\begin{align} 
& \phi_{\alpha} \rightarrow  \Psi^{\dagger D}_1 \sigma_+ \Psi^{D}_2=c^{\dagger }_1 c_2 e^{i \theta_p},~ \phi_{\alpha}\rightarrow \exp(i \int \omega^{xy}_i dx_i)\phi_{\alpha}  \nonumber\\
& \phi_{\beta} \rightarrow  \Psi^{\dagger D}_1 \sigma_- \Psi^{D}_2=c^{\dagger }_1 c_2 e^{-i \theta_p},  ~ \phi_{\beta}\rightarrow \exp(-i \int \omega^{xy}_i dx_i)\phi_{\beta} 
\label{spin}
\end{align}
These are the only excitons with nonzero topological spin ($\pm 1$), else channels has spin zero. $\phi_{\alpha}, \phi_{\beta}$ contains $\pm \pi$ winding number in momentum space.  Such internal structure with nonzero angular momentum dress the exciton with topological spin.
Akin to the Fu-Kane SC, the internal Berry phase of the original Fermi surface is responsible for the $p+ip$ structure of the exciton order parameter.

\begin{figure}[h!]
 \centering
  \includegraphics[width=0.35\textwidth]{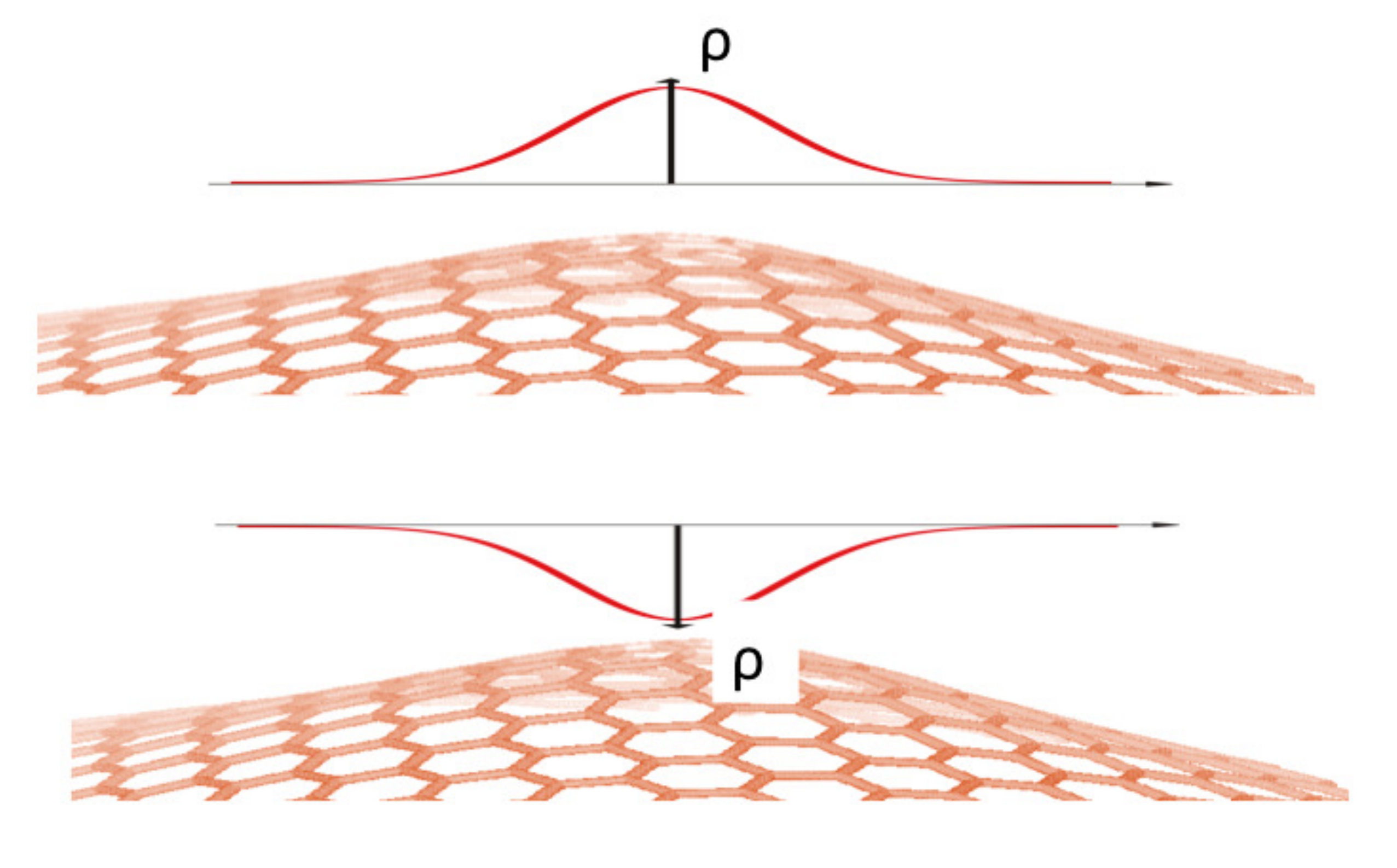}
 \caption{When placing the bilayer half-filled LL on a conic background, the electron density is modulate in opposite ways in different layers, due to the Wen-Zee effect.}
 \label{wenzee}
 \end{figure}

The effective theory for $\phi_{\alpha}$ condensed phase is,
\begin{align} 
&\mathcal{L}_{\phi_{\alpha}}=|(i\partial_{\mu}+2a^-_{\mu}+\omega^{xy}_{\mu})\phi_{\alpha}|^2-s|\phi_{\alpha}|^2-r/2 |\phi_{\alpha}|^4 \nonumber\\
&-\frac{1}{2\pi} A^+ da^+ -\frac{1}{2\pi} A^- da^-+...
 \nonumber\\
&a^{\pm}=\frac{a_1\pm a_2}{2},A^{\pm}=\frac{A_1\pm A_2}{2},
\label{actionalpha}
\end{align}
(Here we omit the damping term $ \frac{i\Omega}{q} |\phi_{\alpha}|^2$ which is nonlocal.)
The exciton condensation does not affect the $a^+$ gauge field. As the system still contains Fermi surface, the gauge field $a^+$ is overdamped  and exhibits metallic behavior. The gauge fields $(2a^-_{\mu}+\omega^{xy}_{\mu})$ would be Higgsed. The finally external response between the spin connection and EM field has the form of Wen-Zee term,
\begin{align} 
&\mathcal{L}=\frac{1}{4\pi} A^- d\omega^{xy}
\label{wz}
\end{align}
This indicates the geometry curvature would modified the electron density on each layer in opposite ways. 
\begin{align} 
&\rho^e_1-\rho^e_2=\frac{1}{8\pi}  d\omega^{xy}=\frac{1}{8\pi}  \frac{\sqrt{g}}{2}R^{xy}
\end{align}
If we put such bilayer system in curved background, the electron charge density would be modified accordingly as Fig.\ref{wenzee}. Due to the nonzero geometry curvature, there should be an electron density imbalance between the two layers. Such phenomenon is not expected for interlayer coherent composite Fermi Liquid phase in HLR side. Even the composite fermion $\Psi^{cf}$ carries topological spin due to flux attachment\cite{Wen_Zee,Cho-2014,Gromov2014,you2016nematic,you2014theory}, the exciton $\Psi^{\dagger,cf}_1\Psi^{cf}_2$ pair involving particle hole channel carries zero topological spin. Hence, such Wen-Zee like response is unique in CDL as a consequence of Dirac nature of the Fermi surface.

In experiment, the Wen-Zee response is recently verified by synthetic Landau levels for photon placing on a conic geometry\cite{schine2016synthetic}. The striking density modulation on the tip of cone is observed as a smoking gun of Wen-Zee effect. In addition, recent numerical simulation\cite{wu2016fractional} also reveals the density modulation for different FQH states on conic geometries. Thus we expect the exciton condensation induced Wen-Zee effect in Eq.(\ref{wz}) could be observed in experiment and numerics to justify Son's CDL theory.

\section{Summary and outlook}
In this work, we explored the interlayer coherence between composite Fermi surface in bilayer Half-filled LL. The p-wave symmetry and topological spin of the exciton order parameter are built from the Berry phase hidden in the composite Fermi surface and provide us with a promising way to differentiate Son's theory from HLR. The Wen-Zee effect and the Lifshitz criticality appeared in our ICCFL transition is unique in Son's CDL theory and the Dirac Fermi surface structure is responsible for these exotic phenomenon.

However, we would like to comment that the nodal structure and topological spin in some specific interlayer coherence channels are merely valid in the limit of PH symmetry and projected Landau Levels. Without PH symmetry, one cannot classify $\phi$ in terms of PH even/odd while the nodal structure is only robust at PH odd channel. In addition, the topological spin is defined with respect to the guiding center geometry after LL projection. The locking between momentum and spinor(dipole) could be damaged once we break PH symmetry and hence the topological spin is ill-defined.  Therefore, the difference between CDL and HLR theory only survives when we have LL projection and PH symmetry. Once we go away from this limit, the distinction eventually disappears.

Apart from interlayer coherent composite fermi liquid state, it is interesting to explore other possible phases and phase transitions in bilayer systems. When the bilayer half-filled Landau Levels are at short distance, the systems goes into an interlayer superfluid (also known as (111)) state $\langle \Psi^{\dagger e}_1  \Psi^{e}_2 \rangle \neq 0$ where the physical electron acquires interlayer coherence. This interlayer superfluid order parameter should be distinguished from the exciton we defined in our previous content as it involves tunneling between physical electrons instead of composite fermions. In the HLR side, the transition between coherent composite fermi liquid state toward (111) state is driven by the condensation of superfluid boson coupling with a dynamical compact U(1) gauge theory, as is discussed in Section II(A) in detail\cite{alicea2009interlayer}.  Such novel quantum criticality contains rich physics and anomalous transport behavior\cite{senthil2008critical}.

In Son's Dirac Fermi surface approach, the (111) state is realized by interlayer pairing of the composite Fermi surface. The interlayer pairing is enhanced by gauge fluctuation of $a^-$ and thus more favorable at short distance. The transition between interlayer coherent composite fermi liquid state and (111) state at this stage cannot be connected by a continuous transition and an intermediate phase with $Z_4$ topological order in between is highly possible\cite{sodemann2016composite}.

\begin{acknowledgements}
We are grateful to J-Y Chen, I Sodemann, I Kimchi, C Wang, S Todadri and E Fradkin for insightful comments and discussions. We also appreciate B Lian for sharing his unpublished work.
 This work was supported in part by the Graduation Dissertation Fellowship (YY) at the University of Illinois. 
\end{acknowledgements}

\appendix
\begin{widetext}
\section{Singularities in Lindhard function}
In this section, we provided calculation detail on density-density correlation at static limit in the ICCFL phase.
\begin{align} 
&\langle \rho^{cf} \rho^{cf} \rangle(q, \Omega=0)=-\frac{\delta S}{\delta a_0 \delta a_0}=
\int d\bm{k}d \omega  ~G(k)G(k+q)\nonumber\\
&G(k)=\frac{1}{\omega-(\epsilon(k)+\Delta)+i\eta}+\frac{1}{\omega-(\epsilon(k)-\Delta)+i\eta} \nonumber\\
&\epsilon(k)=k^2-k_f^2
\end{align}
Here we take $m=1$ and $\Delta$ is related with the strength of the exciton. The Green function of the ICCFL contains two Fermi surfaces with different wave vectors due to interlayer coherence.

We can split the density-density correlation into four independent contributions.
\begin{align} 
&\langle \rho^{cf} \rho^{cf} \rangle(q, \Omega=0)= f_1+f_2+f_3+f_4\nonumber\\
&f_1=\int d \omega d\bm{k} ~\frac{1}{\omega-(\epsilon(k)+\Delta)+i\eta} \frac{1}{\omega-(\epsilon(k+q)+\Delta)+i\eta}\nonumber\\
&f_2=\int d \omega  d\bm{k} ~\frac{1}{\omega-(\epsilon(k)-\Delta)+i\eta} \frac{1}{\omega-(\epsilon(k+q)-\Delta)+i\eta}\nonumber\\
&f_3=\int d \omega  d\bm{k} ~\frac{1}{\omega-(\epsilon(k)+\Delta)+i\eta} \frac{1}{\omega-(\epsilon(k+q)-\Delta)+i\eta}\nonumber\\
&f_4=\int d \omega d\bm{k} ~\frac{1}{\omega-(\epsilon(k)-\Delta)+i\eta} \frac{1}{\omega-(\epsilon(k+q)+\Delta)+i\eta}
\end{align}
Obviously $f_1,f_2$ are the Lindhard function of a Fermi surface with wave vector $\sqrt{k_f^2 \pm \Delta}$. They contain singularities at $2\sqrt{k_f^2 \pm \Delta}$ and the Lindhard function goes as $\langle \rho^{cf} \rho^{cf} \rangle(q, \Omega=0)\sim 1-\Theta(q-2\sqrt{(k_f^2 \pm \Delta)})\sqrt{1-4(k_f^2 \pm \Delta)/q^2}$.

Now we turn to $f_3,f_4$,
\begin{align} 
&f_3=\int d\bm{k} ~\frac{\theta(k-\sqrt{k_f^2 - \Delta})}{-q^2-2kq cos(\theta_k)+2\Delta} +\frac{\theta(|k+q|-\sqrt{k_f^2 + \Delta})}{q^2+2kq cos(\theta_k)-2\Delta}\nonumber\\
&f_4=\int d\bm{k} ~\frac{\theta(|k+q|-\sqrt{k_f^2 - \Delta})}{q^2+2kq cos(\theta_k)+2\Delta} +\frac{\theta(k-\sqrt{k_f^2 + \Delta})}{-q^2-2kq cos(\theta_k)-2\Delta}
\end{align}
Sum over the first terms in $f_3,f_4$, we obtain
\begin{align} 
&\int d\bm{k} ~\frac{\theta(k-\sqrt{k_f^2 - \Delta})}{-q^2-2kq cos(\theta_k)+2\Delta} +\frac{\theta(|k+q|-\sqrt{k_f^2 - \Delta})}{q^2+2kq cos(\theta_k)+2\Delta} \nonumber\\
&=\int^{\sqrt{k_f^2 - \Delta}}_0 dk \frac{k}{q^2-2\Delta} \frac{1}{\sqrt{1-4q^2k^2/(q^2-2\Delta)^2}} \nonumber\\
&=\frac{q^2-2\Delta}{2q^2}(1-\sqrt{1-\frac{4q^2(k_f^2 -\Delta)}{(q^2-2\Delta)^2}}) ~ ~(q<\sqrt{k_f^2 + \Delta}-\sqrt{k_f^2 - \Delta}); \nonumber\\
& \frac{q^2-2\Delta}{2q^2}~~(\sqrt{k_f^2 + \Delta}+\sqrt{k_f^2 - \Delta}>q>\sqrt{k_f^2 + \Delta}-\sqrt{k_f^2 - \Delta}); \nonumber\\
&\frac{q^2-2\Delta}{2q^2}(1-\sqrt{1-\frac{4q^2(k_f^2 -\Delta)}{(q^2-2\Delta)^2}})~~(q>\sqrt{k_f^2 + \Delta}+\sqrt{k_f^2 - \Delta})
\end{align}

Likewise, Sum over the second terms in $f_3,f_4$, we obtain
\begin{align} 
&\int d\bm{k} ~\frac{\theta(|k+p|-\sqrt{k_f^2 + \Delta})}{q^2+2kq cos(\theta_k)-2\Delta} +\frac{\theta(k-\sqrt{k_f^2 + \Delta})}{-q^2-2kq cos(\theta_k)-2\Delta} \nonumber\\
&=\frac{q^2+2\Delta}{2q^2}(1-\sqrt{1-\frac{4q^2(k_f^2 +\Delta)}{(q^2+2\Delta)^2}}) ~ ~(q<\sqrt{k_f^2 + \Delta}-\sqrt{k_f^2 - \Delta}); \nonumber\\
& \frac{q^2+2\Delta}{2q^2}~~(\sqrt{k_f^2 + \Delta}+\sqrt{k_f^2 - \Delta}>q>\sqrt{k_f^2 + \Delta}-\sqrt{k_f^2 - \Delta}); \nonumber\\
&\frac{q^2+2\Delta}{2q^2}(1-\sqrt{1-\frac{4q^2(k_f^2 +\Delta)}{(q^2+2\Delta)^2}})~~(q>\sqrt{k_f^2 + \Delta}+\sqrt{k_f^2 - \Delta})
\end{align}
Thus, the $f_3,f_4$ contributes singularity at $\sqrt{k_f^2 + \Delta}\pm\sqrt{k_f^2 - \Delta}$ to the density-density correlator. Here $\sqrt{k_f^2 + \Delta}-\sqrt{k_f^2 - \Delta}$ is the splitting distance between two Fermi surface while $\sqrt{k_f^2 + \Delta}+\sqrt{k_f^2 - \Delta} \sim 2k_f$ at small splitting.

However, this density density correlator is merely the correlator of the composite fermions. To obtain the EM response measured in experiments, we take the relation $\langle J_x^{e} J_x^{e} \rangle=q_y^2\langle a_0 a_0\rangle$(Based on Son's theory). Hence, the singularity of the Lindhard function for composite Fermi surface would survive in the EM response of Half-filled Landau Levels.

In addition, in this calculation, the energy dispersion of the composite Fermi surface far below the chemical potential is non-relativistic. However, even if we start with relativist fermions, the singularity still survives but the asymptotic behavior might change. Although the emergent Fermi surface in half-filled Landau level is described by the Dirac Fermi surface, the electron far below the Fermi surface is not essentially relativistic\cite{balram2016nature}.

\section{Exciton with internal angular momentum in HLR theory}
For interlayer coherent composite Fermi liquid in HLR theory\cite{alicea2009interlayer,cipri2014gauge}, the exciton order parameter is uniform in space and one does not expect any p-wave structure as long as the interlayer fermion interaction is within the $l=0$ channel.

However, in some circumstances, when appropriate interlayer interaction with higher angular momentum channel is added, the order parameter with respect to the interlayer coherent composite Fermi liquid might carries internal angular momentum $\phi=|\phi| e^{is \theta_p}$. Hence, the exciton order parameter $\phi$ carries topological spin $s$. The corresponding effective theory of the ICCFL is,
\begin{align} 
&\mathcal{L}_{\phi}=|(i\partial_{\mu}+2a^-_{\mu}+s \omega^{xy}_{\mu})\phi|^2-s|\phi|^2-r/2 |\phi|^4 \nonumber\\
&+\frac{1}{4\pi} (A^+ +a^+)\wedge  d(A^+ +a^+) + \frac{1}{4\pi} (A^- +a^-) \wedge d (A^- +a^-)+...
\end{align}
The exciton condensation does not affect the $a^+$ gauge field and the system still maintains metallic behavior. The gauge fields $(2a^-_{\mu}+s\omega^{xy}_{\mu})$ would be Higgsed. The finally mixing response between the spin connection and EM field has the form,
\begin{align} 
&\mathcal{L}=\frac{1}{8\pi} (A^- +s\omega^{xy}_{\mu}) \wedge d (A^- +s\omega^{xy}_{\mu}) 
\end{align}
This theory also contains Wen-Zee response due to the topological spin carried by exciton, akin to the result we got for ICCFL in Son's theory. However, it also contains additional Gravitational Chern-Simons term which is responsible for orbital spin variance. This term is not presented in Son's side so one can still distinguish between HLR and Son's theory. 
\end{widetext}
\providecommand{\noopsort}[1]{}\providecommand{\singleletter}[1]{#1}%


\begin{thebibliography}{68}
\bibitem [{aaa()}]{aaa}%
  \BibitemOpen
  \href@noop {} {}\bibinfo {howpublished} {In order to well defined PH symmetry
  for exciton, we have to claify how layer switch symmetry first. Otherwise,
  one can always add a phase to the exciton and the PH symmetry
  changed.}\BibitemShut {Stop}%
\bibitem [{bbb()}]{bbb}%
  \BibitemOpen
  \href@noop {} {}\bibinfo {howpublished} {We do not consider the mixing
  exciton channel between $\phi_a$ and $\phi_b$ which mixed $\mathcal{CP}$
  even/odd channel.}\BibitemShut {Stop}%
\end{thebibliography}
\end{document}